\documentclass[prl,twocolumn,showpacs,preprintnumbers,amsmath,amssymb]{revtex4}

\usepackage{graphicx}

\newcommand{\pow}{\mathop{\textrm{pow}}}
\newcommand{\opt}{{\mathop{\mathrm{opt}}}}

\newcommand{\Y}{Y}

\begin{document}
\author{A. G. Rossberg}
\email{axel@rossberg.net} 
\affiliation{Zentrum f{\"u}r Datenanalyse und Modellbindung,
Universit{\"a}t Freiburg, Eckerstr.~1, 79104 Freiburg, Germany}
\date{August 5, 2003}
\title{A generic scheme for choosing models and characterizations of
  complex systems}



\begin{abstract} 
  It is argued that the two problems of choosing characterizations and
  models of complex systems should not be considered independently.  A
  particular criterion for these choices, oriented on the potential
  usefulness of the results, is considered, and a generic
  formalization applicable to realistic experiments is developed.  It
  is applied to Kuramoto-Sivashinsky chaos.
\end{abstract}

\maketitle

The systematic characterization of self-organized structures is a
long-standing challange to the science of structure formation.
`Labyrinths', `breathers', `dendrites', `worms', `spiral-defect
chaos', or `scale-free networks' are only few of the words that were
introduced to describe real and numerical experimental observations.
Images and natural language can usually communicate what is ment, but
as the number of observed structures increases and distinctions become
finer, a more systematic approach seems desirable.  The problem is
felt particularly strong for the large variety of spatially irregular
structures and spatio-temporally chaotic states that have been found
\cite{crohop}.

In search for appropriate characterizations researchers do often
concentrate on those properties of the experimental data that are
easily modeled -- those properties of the data or the underlying
structures that are governed by their own rules (one might call them
``coherent structures'').  In this case, \emph{the choice of the
  characterization depends on the available models.}
On the other hand, only when a particular set of properties of
experimental data has been found to be characteristic for an observed
structure, one can meaningfully ask for a model that reproduces this
structure, i.e., a model that reproduces data with these properties.
\emph{Modeling requires prior characterization.}

Intuition is the fallback most researchers rely on when facing this
circular relationship of modeling and characterizing.  In fact,
intuition is an excellent guide.  But for some problem areas, e.g.,
those involving spatio-temporal chaos, progress appears to have slowed
down also due to a lack of intuition about what the characteristic
properties and what appropriate models are.  Even when intuition
\emph{is} successful in choosing models and characterizations, it is
legitimate to ask if these choices are subjective in the sense that
they depend essentially on the way humans observe the world (other
beings might decide very differently), or if they are the solution of
some objective problem, that our intuition is just highly efficient in
solving.  Most of the approaches to the related problem of emergence
(e.g.,
\cite{crutchfield94:_calcul_emerg,baas94:_emerg_hierar_hyperSHORT})
are based on the \textit{a priori} assumption of some limitation to
observation (coarse graining), thus involving an ``inherently
subjective'' \cite{crutchfield94:_calcul_emerg} component.  For an
argument in favor of the objectivity of the choices it is therefore
important to formulate a criterion that does not depend on such
artificial limitations.



Here, a proposal for such a criterion is introduced.  It is first
stated on a heuristic level and then modeled in a mathematical
language; thus modeling the problem of modeling.  This involves the
combination of concepts from computer science that proved powerful in
the context of structure formation -- algorithmic complexity (program
length)
\cite{machta93:_comput_compl_patter_format,%
moore00:_inter_diffus_limit_aggreg%
} and computational complexity (execution time) \cite{dewey97:_algor}
-- with ideas from statistical test theory
\cite{lehmann97:_testinSHORT}.  It is shown that the circular relation
between models and characterizations is, in this case, not vicious:
the criterion leads to nontrivial choices.  As an example, the
formalism is applied to the spatio-temporally chaotic solutions of the
Kuramoto-Sivashinsky Equation.


Consider the following requirements for models and characterizations:
\newlength{\quotewidth} \setlength{\quotewidth}{\columnwidth}
\addtolength{\quotewidth}{-5em}
\begin{align}
  \label{char_cond}
  \begin{minipage}[c]{\quotewidth}
    \emph{Characterizations} should be easily communicated and
    verified, be specific, and should, over a wide control-parameter
    range, apply to experimental data and be reproducible in models.
  \end{minipage}
  \\[1ex]
  \label{mod_cond}
  \begin{minipage}[c]{\quotewidth}
    \emph{Models} should be easily communicated and easily evaluated,
    show little artifacts, and reproduce given characterizations.
\end{minipage}
\end{align}
The practical relevance of most of these requirements is obvious.  To
see why it is desirable that characterizations are reproducible in
models, notice that, from such models, larger, composed models could
be constructed, that can then be used to explore and characterize
situations not accessible experimentally (e.g., climate models).  Even
though the existence of models of sub-systems that reproduce the
properties relevant for the composed model is not guaranteed, in case
that they exist, it is good to know them.
Now, as the general criterion, choose those pairs of models and
characterizations that jointly statisfy conditions~(\ref{char_cond})
and~(\ref{mod_cond}) as well as possible.

\begin{figure}[t]
  \centering
  \includegraphics[width=1.0\columnwidth,keepaspectratio]{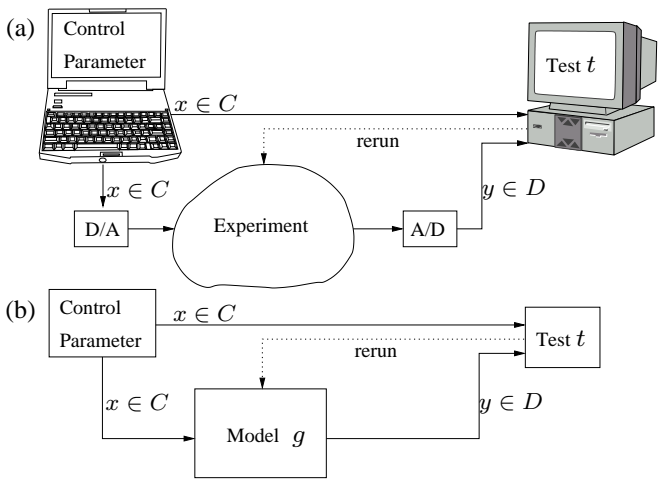}
  \caption{(a)  Generic setup of a computer-controlled experiment.
    (b) Data flow in a test of a computational model.}
  \label{fig:setting}
\end{figure}

In order to formalize this criterion and make it accessible to a
rigorous analysis, both characterizations and models are represented
by computer programs: programs that test data for particular
properties, and programs that generate data.  The practical use of
these programs is illustrated in Fig.~\ref{fig:setting}.

Figure~\ref{fig:setting}a shows a generic setup for a
computer-controlled experiment.  The experimenter enters some control
parameter values at a console.  The set of control parameter values is
encoded in a binary string $x\in C$, where the \emph{control parameter
  format} $C$ is a subset of the set $\{0,1\}^n$ of all binary strings
of length $n$ for some $n\in\mathbb{N}_0$.  Based on these control
parameter values, the control parameters of the experiment are
adjusted, usually \textit{via} some D/A conversion.

When the experiment is run, experimental data is recorded in binary
form.  Data is encoded in a binary string $y\in D$, where the
\emph{data format} $D$ is a subset of the set $\{0,1\}^m$ of all
binary strings of length $m$ for some $m\in\mathbb{N}_0$.  This could,
for example, be image data or a time series.  Generally, $y$ is a
realization of a random variable $Y$ with values in $D$.  The
experiment is assumed reproducible in the sense that repeated runs of
the experiment yield a sequence $Y_1,Y_2,\ldots$ of statistically
independent, identically distributed (i.i.d.)\ results.

A characterization is represented by a program $t$ that computes a
statistical test on experimental data: A \emph{test} takes a control
parameter $x\in C$ as input, runs, and then either halts with output
$\mathbf{e}$ or requests a finite number of (re)runs of the experiment
and then halts with an output of $1$ or $0$.  By the output
$\mathbf{e}$ the tests $t$ indicates that $x$ is not within its range
of validity $C[t]:=\{x\in C|\text{output of $t$ with input $x$ is not
  $\mathbf{e}$\}}$ \footnote{This suppresses simple characterizations
  that are valid only in complex subsets of $C$.}.  The outputs $1$ or
$0$ indicate that the null hypothesis (see below) is accepted or
rejected by the test, respectively.  When the test requests an
experimental rerun, its execution is suspended until the experimental
result $y$ is written into a dedicated storage accessible by the test.

A model is represented by a computer program $g$ that generates data
to be used in place of experimental data (Fig.~\ref{fig:setting}b): A
\emph{generator} takes a control parameter $x\in C$ as input, runs,
outputs data $y\in D$ and halts.  In order to produce random results,
the program has access to a source of independent, evenly distributed
random bits.  Subsequent runs of a generator are fully independent.

As in conventional statistical test theory
\cite{lehmann97:_testinSHORT}, the \emph{power function} is
introduced.  Denote by $t_x(\{y_i\})$ the output of the test $t$ at
control parameter $x\in C[t]$ when applied to the sequence of
experimental results $\{y_i\}\in D^\infty$ (for formal simplicity, the
sequences $\{y_i\}$ are assumed infinite, even though the tests use
only finite subsequences).  Let $\{Y_i\}$ be a sequence of i.i.d.\ 
random results with values in $D^\infty$.  Define the power of the
test function $t_x$ when applied to $\{Y_i\}$ as the probability to
reject $\{Y_i\}$, i.e.,
\begin{align}
  \label{def:power}
  \pow(t_x,\{Y_i\}):=\Pr\!\left[t_x(\{Y_i\})=0 \right]\quad(x\in C[t]).
\end{align}

Unlike in conventional test theory, there is no independent null
hypothesis $H_0$ here that states the distribution or the class of
distributions of $\{Y_i\}$ that is tested for.  Instead, given a test
function $t_x$, the null hypothesis, i.e., the class of distributions,
is \emph{defined} by the condition
\begin{align}
  \label{def:correct_tests}
  \pow(t_x,\{Y_i\})\le\alpha,
\end{align}
where $0<\alpha<1$ is a fixed significance level
\cite{lehmann97:_testinSHORT} \footnote{From $t_x$, computable tests
  for the same $H_0$ at other significance levels can be
  constructed.}.

The ease or difficulty of communicating a test $t$ or model $g$,
mentioned in requirements~(\ref{char_cond},\ref{mod_cond}), is
measured by the lengths $L(t)$, $L(g)$ of the programs $t$ and $g$.
The value of $L(\cdot)$ depends on the machine model.  In the example
below, MMIX, an idealized modern microprocessor is used
\cite{knuth99:_mmixw}.

The ease or difficulty of verifying characterizations and evaluating
models is measured by the execution times $T(g)$, $T(t)$ of the
programs.  To be specific, define $T(\cdot)$ as the maximum of the
expectation value of the runtime over all $x\in C$ and all
distributions of data.  Below, time is measured by the number of
``oops'' (symbol: $1\,\upsilon$) counted by the MMIX emulation
\texttt{mmix-sim} \cite{knuth99:_mmixw}.

The often-encountered tradeoff between $L$ and $T$ is taken into
account by assuming that there is a cost function depending on both
resources, which increases strictly monotonically with $L$ at fixed
$T$ and with $T$ at fixed $L$ but is otherwise unspecified.  With
this in mind, define the relations $\preceq$ (\textit{always cheaper
  or equal}) and $\prec$ (\textit{always cheaper}) for programs $p_1$,
$p_2$ by
\begin{align}
  \label{def:cheaperorequal}
  p_1 \preceq p_2 &\stackrel{\text{def}}{\Leftrightarrow}
  \text{$L(p_1)\le L(p_2)$ and $T(p_1)\le T(p_2)$}\\
  \intertext{and}
  \label{def:cheaper}
  p_1 \prec p_2 &\stackrel{\text{def}}{\Leftrightarrow}
  \text{$p_1\preceq p_2$ and not $p_2\preceq p_1$}.
\end{align}
%
It turns out that the machine dependence of relations $\preceq$ and
$\prec$ for implementations of algorithms on different processor
models is weak.
In principle, other resources could also be taken into account in
definition~(\ref{def:cheaperorequal}) such as, for tests, the number
of experimental runs required.

Since for every program $p$ there is only a finite number of programs
with smaller or equal length, there is also only a finite number of
programs $p'$ such that $p'\prec p$ or $p'\preceq p$.  Below we need
\textit{Lemma 1:} \textit{Every nonempty set $P$ of tests or
  generators contains an element $p$ which is minimal with respect to
  the relation $\prec$, i.e., such that no $p'\in P$ satisfies
  $p'\prec p$.}  This is a direct consequence of the previous note and
the transitivity and antireflexivity of $\prec$.  In general, there
are several minimal elements, each using its own mix of resources.
This reflects the intuition that there are several ``good'' models and
characterizations for one experiment.

These concepts from statistics and computer science are now combined
to formalize requirement~(\ref{mod_cond}), except for the condition
regarding artifacts.
Denote by $g_x$ the sequence $\{Y_i\}$ of random outputs of generator
$g$ at control parameter $x$.  Define for given $C$, $D$ the notion of
an \emph{optimal generator} $g$ relative to a test $t$ and a power
threshold $1>\gamma>\alpha$ by
\begin{subequations}
  \label{def:opt}
  \begin{align}
    \label{def:opt_sat}
    \opt_t^\gamma g\stackrel{\text{def}}{\Leftrightarrow} &
    \bigwedge_{x\in C[t]} \pow(t_x,g_x) \le \alpha \text{ and}\\
    \label{def:opt_opt}
    &\bigwedge_{g'\prec g}\,\bigvee_{x\in C[t]}
    \pow(t_x,g_x')>\gamma\text{ and}\\
    \label{def:opt_eql}
    &\bigwedge_{g'\preceq g}\,\bigvee_{x\in C[t]} \pow(t_x,g_x') \ge
    \pow(t_x,g_x),
  \end{align}
\end{subequations}
where the quantifiers $\bigwedge$ (\textit{for all}) and $\bigvee$
(\textit{there is}) have been introduced for brevity.
Line~(\ref{def:opt_sat}) states that $g$ satisfies $t$,
line~(\ref{def:opt_opt}) says that all cheaper generators are rejected
by $t$ with power $>\gamma$ and line~(\ref{def:opt_eql}) handles the
generators that use the same resources as $g$. 
The test $t$ is specific to $g$ in the sense that it does not apply to
any $g'\prec g$.

In order to disentangle the circularity between models and
characterizations, consider now the problem of specifying a generator
by characterizing its output.  For a i.i.d.\ random sequence $\{Y_i\}$
denote by $p[\{\Y_i\}]$ the distribution function of its elements,
i.e., $p[\{Y_i\}](y):=\Pr[Y_1=y]$ for $y\in D$.  Call a generator $g$
an \emph{optimal implementation} with respect to a set $\tilde
C\subset C$ iff there is no generator $g'\prec g$ such that
$p[g'_x]\equiv p[g_x]$ for all $x\in \tilde C$.  \textit{Theorem 1}:
\textit{Given $C$ and $D$, there is for every $\tilde C\subset C$,
  every optimal implementation $g$ with respect to $\tilde C$, and
  every $1>\gamma>\alpha$, a test $t$ such that $\opt_t^\gamma g$ and
  $C[t]=\tilde C$.} \emph{Outline of the proof:} Explicitly construct $t$.
$x\in\tilde C$ can be tested for by keeping a list of $\tilde C$ in
$t$.  Since there is only a finite number of $g'\preceq g$, the test
must distinguish $p[g_x]$ from a finite number of different
distributions $p[g'_x]$ for all $x\in\tilde C$, with certainty
$\gamma$ if $g'\prec g$.  This can be achieved by comparing a
sufficiently accurate representation of $p[g_x]$, stored in $t$ for
all $x\in \tilde C$, with a histogram sampled from $g'_x$.  With a
high number of samples, any degree of certainty can be reached.$\Box$
The cost of testing is not taken into account, yet.

The following definition formalizes the criterion stated above for
choosing models and characterizations; to find pairs $(t,g)$ jointly
satisfying conditions~(\ref{char_cond},\ref{mod_cond}) as well as
possible.  Only the validity of characterizations for experiments is
not contained in the definition:
Given $C$ and $D$, call a pair $(t,g)$ a \emph{basic model specifying
  characterization} (\textit{b.m.s.c.})\ iff there is a
$1>\gamma>\alpha$ such that $\opt^\gamma_t g$ and there is no $t'\prec
t$ with $C[t]\subset C[t']$ and $\opt^\gamma_{t'} g$.

This optimization with respect to $t$ implies the avoidance of
artifacts, when artifacts are considered as properties that are
specific and are cheaper to communicate and verify than the property
$t$ that $g$ is supposed to model.
The definition of a \textit{b.m.s.c.}\ involves the simultaneous
minimization of cost with respect to $t$ and $g$.  An answer to the
question if there are any nontrivial solutions to this double
optimization problem -- i.e., if the circular relation between models
and characterizations as considered here is vicious -- is given by 
\textit{Theorem~2}: \textit{Given $C$ and $D$, there is, for every
  $\tilde C\subset C$ and every optimal implementation $g$ with
  respect to $\tilde C$, a test $t$ such that $(t,g)$ is a b.m.s.c.\ 
  and $\tilde C \subset C[t]$.}
\emph{Proof:} Fix some $1>\gamma>\alpha$.  By Theorem 1,
the set $S:=\{t|\opt^\gamma_{t} g \text{ and } \tilde C \subset
C[t]\}$ is nonempty.  Theorem~2 is satisfied by any $t\in S$
which is minimal with respect to the half ordering $\prec$.  By
Lemma~1 such an element exists.$\Box$

Only for a few \textit{b.m.s.c.}\ $(t,g)$ the test $t$ also applies to
a given experiment.  Generally, there will be some fundamental level
of description (the Schr{\"o}dinger equation, say) at which a 1-to-1
model $g$ of the experiment can be constructed, and then a
corresponding $t$ exists by Theorem~2.  But these \textit{b.m.s.c.}\ 
are often too expensive.  Finding cheaper \textit{b.m.s.c.}\ that
apply to the data requires intuition, insight, and experience, and
goes beyond the scope of this work.  The goal here was only to
investigate if an objective, well-posed problem of modeling and
characterizing exists, and to model it so that among several solutions
conceived some are selected.

As an example for an application, assume some idealized experiment, the
fundamental description of which is given by the Kuramoto-Sivashinsky
(KS)
equation
\begin{align}
  \label{KS-Equation-u}
  \partial_{\tau} u=-\partial_{\xi}^2 u-\partial_{\xi}^4 u+ u \partial_\xi u,
\end{align}
with periodic boundary conditions $u(\tau,\xi)=u(\tau,\xi+\Xi)$, as
they apply for experiments in a ring-channel geometry.  In each
experimental run, $128$ equally spaced points of $u$ at distance
$\Delta \xi=0.8$ ($\Xi=128\times \Delta \xi$) are sampled $200$ times
in $\Delta\tau=0.2$ sampling intervals, while $u$ is evolving along
the chaotic KS attractor.  The data format $D$ is given by all
sequences of $128\times200=25600$ 8-byte floating point numbers.
There is no control parameter: $x$ is always the empty string and the
only element of $C$.

The systematic construction of a pair $(t,g)$ likely to be a
\textit{b.m.s.c.}\ of the experiment goes from the generator $g$ over
a corresponding test $t$ to a verification that the experiment passes
the test $t$.  Practically finding a suitable $g$ requires a
preliminary approximation of $t$ characterizing the experiment.  This
first, exploratory step is not described here.

\begin{figure}[t]
  \centering
  \begin{minipage}[b]{0.3\columnwidth} 
    \flushleft a)\\
    \includegraphics[width=\columnwidth,keepaspectratio]{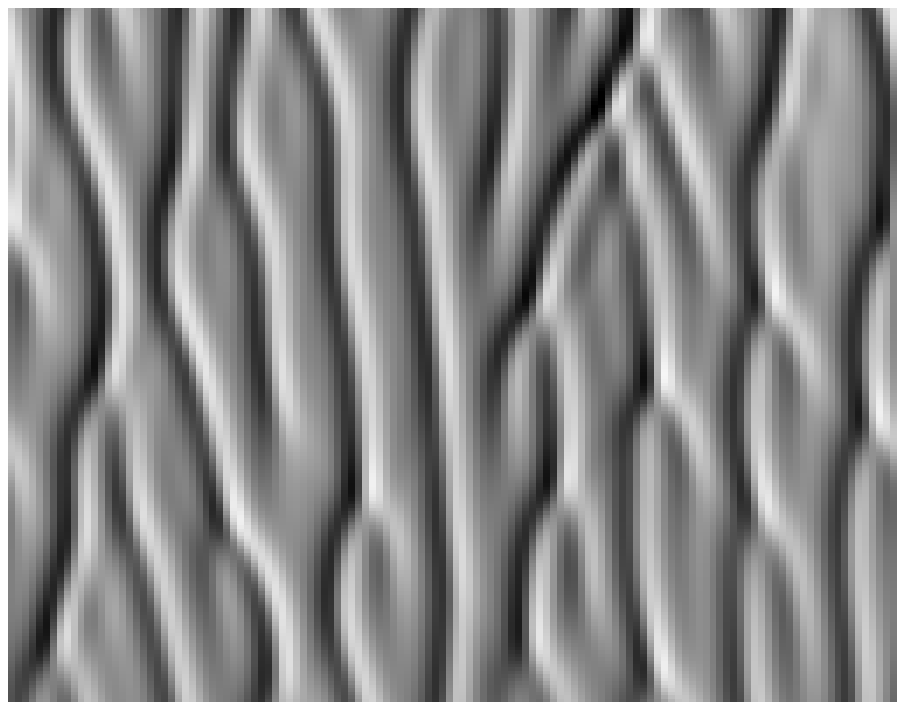}
  \end{minipage}
  \begin{minipage}[b]{0.3\columnwidth} 
    \flushleft b)\\
    \includegraphics[width=\columnwidth,keepaspectratio]{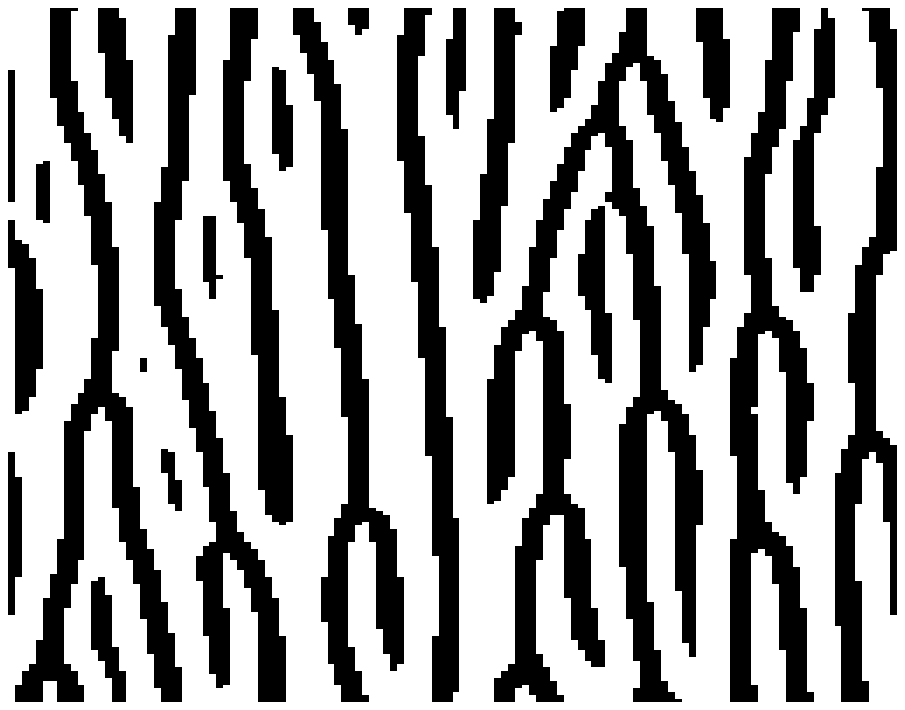}
  \end{minipage}
  \begin{minipage}[b]{0.3\columnwidth} 
    \flushleft c)\\
    \includegraphics[width=\columnwidth,keepaspectratio]{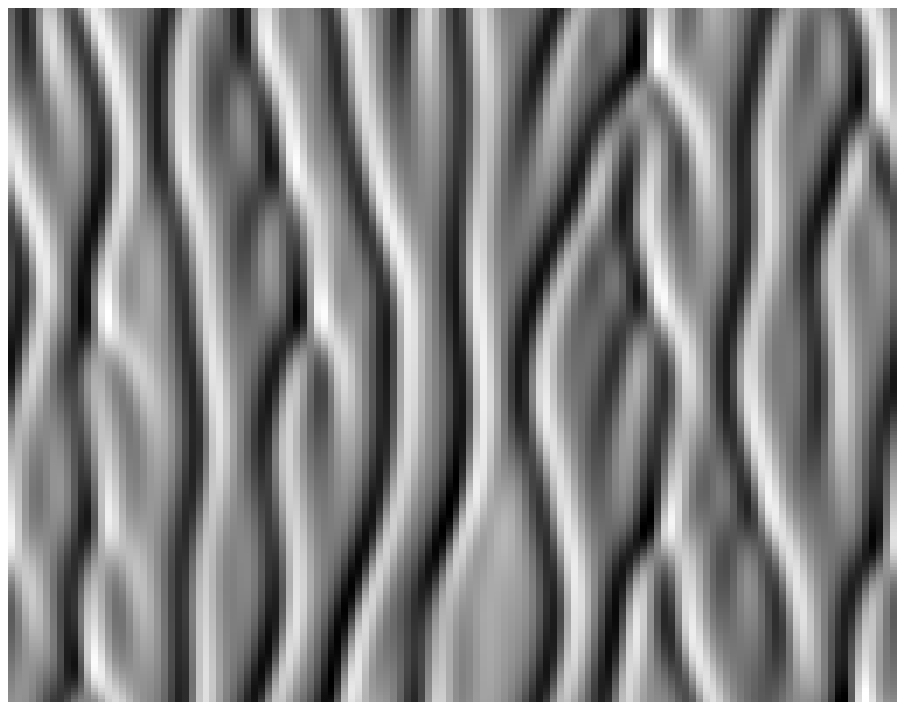}
  \end{minipage}
  \includegraphics[width=0.06\columnwidth,keepaspectratio]{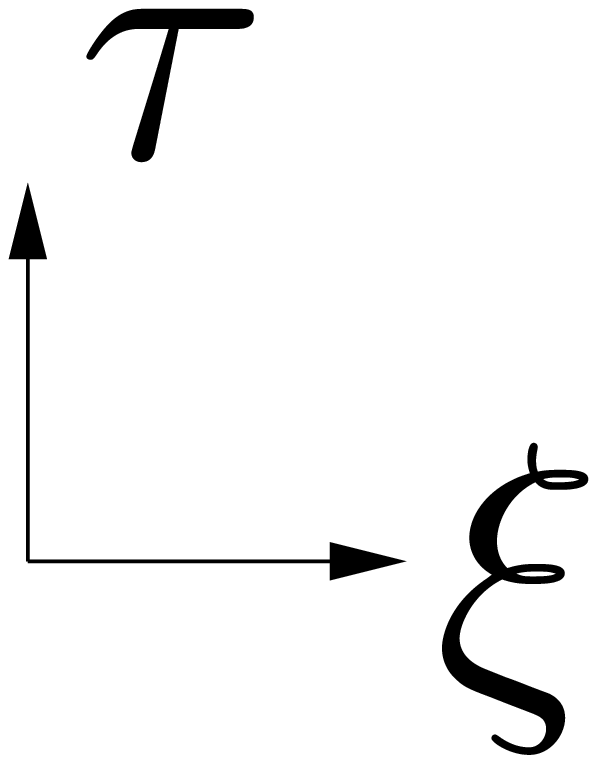}
  \caption{Modeling and characterization of KS
    chaos. (a) Gray coding of the output $y$ of model $g$ that
    approximates Eq.~(\ref{KS-Equation-u}).  (b) Regions where
    $\partial u(\tau,\xi)/\partial \xi>0$ (black) as used in the
    ``tree-test'' $t$. (c) A precise solution of
    Eq.~(\ref{KS-Equation-u}) for comparison.}
  \label{fig:tree}
\end{figure}

The code for the generator $g$ is a minimal-length implementation of
Eq.~(\ref{KS-Equation-u}) on an MMIX processor.  A discretization
$u_{p,q}$ locally approximately proportional to a solution
$u(10\,p\,\Delta \tau ,q\,\Delta x)$ of Eq.~(\ref{KS-Equation-u}) is
obtained by an Euler integration with in-place update of the form
\begin{multline}
  \label{euler-scheme}
  u_{p+1,q}=c_1\,(u_{p,q+2}+u_{p+1,q-2})
  +c_2\,(u_{p,q+1}+u_{p+1,q-1})\\
  +(c_3+u_{p,q+1}-u_{p+1,q-1})\,u_{p,q},
\end{multline}
where $(c_1,c_2,c_3) \approx (-0.05,0.18,0.75)$.  Including code to
handel the periodic boundaries, to initialize $u_{0,\xi}$ with random
numbers $\mathcal{O}(10^{-2})$, to drop a transient of 16 time units,
and to output $y$ (Fig.~\ref{fig:tree}a), this requires
$L(g)=260\,\mathrm{bytes}$ and $T(g)=34\,\mathrm{M\upsilon}$ for a
single run.

For the test $t$, a code is used that implicitly computes the stripes
$\partial_\xi u(\tau,\xi)>0$ (Fig.~\ref{fig:tree}b) using data of
every $k$-th sampling interval ($k\approx 5$). Then it determines for
each of $N=20$ runs the total numbers of beginnings $n_b$, ends
$n_e$, mergers $n_m$, and splits $n_s$ of the stripes along the time
axis, as well as the average number $l$ of stripes.  The value of $N$
implicitly determines $\delta$.

\enlargethispage{2ex}

If a combination $(n_b,n_e,n_m,n_s,l)$ is repeated for two runs, the
test rejects the data stream in order to enforce randomness.  The
averages
$\overline{n}_b,\overline{n}_e,\overline{n}_m,\overline{n}_s,\overline{l}$
of these statistics over all $N$ runs are determined.  The data is
rejected if $\overline{n}_e > m_e\approx 15.2$ or $\overline{n}_s >
m_s\approx 0.4$, which enforces the tree-like geometry of the stripes
and consequently a minimal accuracy of $g$.  Data is rejected if
$(\overline{n}_l-m_l)^2> v_l$ or $(\overline{n}_b-m_b)^2> v_b$, which
sets the length and time scales of the tree structure
$[(m_l,v_l,m_b,v_b)\approx (14.,0.03,22.,1.5)]$.  Finally, data is
rejected if the difference between the initial and final number of
stripes is large, i.e.,
$(\overline{n}_e+\overline{n}_m-\overline{n}_b-\overline{n}_s)^2> v_a
\approx 1.7$, which enforces the suppression of a transient in the
generator.  Within the statistical error, $t$ accepts $g$ at the
$\alpha=0.1$ significance level: $\pow(t_x,g_x)=0.105(3)\lesssim
\alpha$.  Using precise numerical simulations of
Eq.~(\ref{KS-Equation-u}), it was verified that solutions of the
fundamental description (Fig.~\ref{fig:tree}c) are rejected by $t$
with a probability of only $0.03(1)<\alpha$.  That is, $t$
characterizes the ``experimental'' data and is even robust to small
deviations from the fundamental description~(\ref{KS-Equation-u}).
A compiler-optimized implementation \cite{nilsson01:_gcc_mmix_abi} of
$t$ requires $L(t)=1192\,\mathrm{bytes}$ and
$T(t)=3.8\,\mathrm{M\upsilon}=N\times 0.19\,\mathrm{M\upsilon}$.

In principle, the precise values of the tuning parameters in $t$ could
be determined by locally solving the optimization problem for the
condition for the pair $(t,g)$ to be a \textit{b.m.s.c.}\ to the
precision of the coding of the parameters.
Regarding the question if this pair is also a global solution of the
optimization problem for a \textit{b.m.s.c.}, it can only be said that
this is a plausible conjecture.  It has been checked that the direct
verification of Eq.~(\ref{euler-scheme}) would yield a test that is
shorter than the tree-test $t$, but requires much more time.
Likewise, generators more explicitly coded to generate tree structures
accepted by $t$ could be faster than $g$, but the examples
investigated indicate that, due to several conditional branches, they
would always be longer.  Thus, no counterexamples could be found.
Notice that the information reduction performed by $t$ in
concentrating on the stripes $\partial_\xi u>0$ is not externally
imposed.  Rather, it is the a consequence of the rather small number
of competing generators to be excluded.

To the degree that the pair $(t,g)$ described here it is a
\textit{b.m.s.c.}, it is also of practical relevance.  The tree-test
$t$ provides a fast, rather simple, and robust way to identify
KS chaos.  There seems to be no other simple ``explanation'' for the
structure identified by $t$.  On the other hand, $g$ provides a simple
and, as it turns out, comperatively fast method to obtain
approximations of KS chaos on digital computers, which is important
whenever resources are scarce.

A formal scheme combining computation and statistics for choosing
models and characterizations has been laid out.  It models the main
aspects of the practical problem.  The question if the choices are
``intuitive'' is presumably hard to answer systematically.  At least,
it has been argued, they are useful: not because nature is a computer,
but because people use computers.

Work supported by the German BMBF (13N7955).

\end{document}